\documentclass[aps,preprint,showpacs,pre]{revtex4-1}

\usepackage{graphicx}
\usepackage{amsmath}
\usepackage{amssymb}
\usepackage{mathrsfs}
\usepackage{color}
\usepackage{ulem}
\usepackage[utf8]{inputenc}
\usepackage{subfigure,wrapfig}

\newcommand{\be}{\begin{equation}}
\newcommand{\ee}{\end{equation}}
\newcommand{\ba}{\begin{align}}
\newcommand{\ea}{\end{align}}

\begin{document}


\title{A maximum entropy approach to  H-theory: Statistical mechanics  of hierarchical systems}


\author{Giovani L.~Vasconcelos$^1$, Domingos S.~P.~Salazar$ ^2$, A. M. S. Mac\^edo$^1$ }
\affiliation{$^1$ Laborat\'orio de F\'{\i}sica Te\'orica e Computacional, Departamento de F\'{\i}sica, Universidade Federal de Pernambuco 50670-901 Recife, Pernambuco, Brazil}
\affiliation{$^2$Unidade de Educa\c{c}\~ao a Dist\^ancia e Tecnologia, Universidade Federal Rural de Pernambuco, 52171-900
Recife, PE, Brazil}

\email{giovani.vasconcelos@ufpe.br}

\begin{abstract}
A novel  formalism, called  H-theory, is applied to the problem of statistical equilibrium of a  hierarchical complex system with multiple time and length scales. In this approach, the system is  formally treated  as being composed of a small subsystem---representing the region where the measurements are  made---in contact with a set of `nested heat reservoirs' corresponding to the hierarchical structure of the system. The probability distribution function (pdf) of the fluctuating temperatures at each reservoir, conditioned on the temperature of the reservoir above it,  is determined from a maximum entropy principle subject to appropriate constraints that describe the thermal equilibrium properties of the system. The  marginal temperature distribution of the innermost reservoir  is  obtained by integrating  over the conditional distributions of all larger scales, and the resulting pdf is written in analytical form in terms of  certain special transcendental functions, known as the Fox $H$-functions.  The  distribution of states of the small subsystem is then computed by averaging the  quasi-equilibrium Boltzmann distribution over  the temperature of the innermost reservoir. This distribution can also be written in terms of $H$-functions. The general family of distributions reported here recovers, as  particular cases, the stationary distributions recently obtained by Mac\^edo {\it et al.} [Phys.~Rev.~E {\bf 95}, 032315 (2017)] from a stochastic dynamical approach to the problem.
\end{abstract}


\maketitle

\section{INTRODUCTION}

Complex systems with multiple time and length scales occur frequently in many areas of physics and interdisciplinary fields, such as turbulence \cite{frisch},   random-matrix theory \cite{mehta},  high-energy collision physics \cite{wilk,cosmicrays}, and econophysics \cite{finance}, to mention only a few. 
One common feature among many such systems is the appearance  of probability distributions that deviate considerably from what one would expect (say, Gaussian or exponential behavior) on the basis of standard equilibrium statistical mechanics arguments. 
A great deal of effort has therefore been devoted to constructing physical models that generate such heavy-tailed  distributions. One approach that has attracted considerable attention is the so-called nonextensive statistical mechanics formalism \cite{tsallis} whereby a power-law distribution, known as the Tsallis distribution, is obtained by maximizing a nonextensive entropy that generalizes the Boltzmann entropy formula.  
Heavy-tailed distributions can also be accounted for by a superposition of two statistics---a procedure known in mathematics as compounding \cite{compounding} and in physics as superstatistics \cite{beck}. 
In particular, the Tsallis distribution can be readily obtained from the superstatistics approach by  an appropriate choice of the weighting distribution \cite{beck}. Furthermore, this choice of weighting distribution can  be justified from both a Bayesian analysis \cite{sattin,RCF}  and a maximum entropy principle based  on the  Boltzmann-Shannon entropy \cite{VakarinPRE2006,beckpre2007,Crooks2007, BeckPRE08,Dixit}, thus circumventing the need to introduce a non-extensive entropy to justify the emergence of heavy-tailed  distributions.

Recently, we introduced a  general formalism  \cite{SV1, SV2,ourPRE2017} that extends the superstatistics approach to multiscale systems  and gives rise to a large family  of heavy-tailed distributions  labeled by  the number $N$ of  different  scales present in the system. (Usual superstatistics corresponds to $N=1$ \cite{Sobyanin2011}.) In this hierarchical formalism, to which we refer as {\it H-theory}, it is assumed  that at large scales the statistics of the system is described by a known distribution that contains a  parameter (say, the  temperature $T_0$)  that characterizes the global equilibrium of the system.  At short scales, however, the system  deviates considerably from the large-scale distribution, owing to the 
complex  multiscale dynamics (intermittency effects) of the system. 
The scale  dependence of the relevant distributions can be  effectively described by assuming that the environment (background) surrounding  the  system under investigation  changes slowly in time.   The dynamics of the  background  is then  formulated as a set of hierarchical stochastic differential equations whose form is derived from simple physical constraints, yielding  only two `universality classes' for  the stationary distributions of the background variables at each level of the hierarchy: i) a  gamma distribution   and ii) an inverse-gamma distribution.  For both classes, analytical expressions are  obtained for the marginal  distribution of the background variable at the lowest level of the hierarchy  in terms of Meijer $G$-functions, from which the heavy-tailed distribution of the fluctuating signal  is computed (and also written in terms of $G$-functions). Here two classes of signal distributions are found \cite{ourPRE2017} according to the behavior at the tails: i) power-law decay and ii) stretched-exponential tail. Applications of the H-theory to empirical data from several systems, such as turbulence \cite{SV1,SV2}, financial markets \cite{ourPRE2017}, and random fiber lasers \cite{NatComm2017} have yielded  excellent  results.

The  dynamical formulation of the H-theory reviewed in the preceding paragraph represents a
`microscopic'  (i.e., small-scale) approach to the problem,  in that it tries to model the  fluctuations in the environment under which the system evolves  by a set of stochastic differential equations, which  in principle provides a full description of the time-dependent stationary joint distribution function of the background variables. In this paper we take  an alternative,   thermodynamic-like approach in which  the  background distribution will be derived from a maximum entropy principle, thus bypassing the need to specify the underlying dynamics. We remark that this weakening of the basic dynamical hypothesis of H-theory leads to a considerable expansion of its domain of applicability, which may now include complex multiscale systems with non-Markovian stochastic dynamics.  

The main purpose of the paper is  to present a unified maximum-entropy principle suitable for hierarchical complex systems in statistical equilibrium. The main idea in our approach is to write the  Boltzmann-Shannon  entropy of the system  in terms of the local equilibrium distribution of states
and  the distributions of the background variables (`local temperatures') across the hierarchy.   In other words, the system is treated  as  being  effectively composed of a small system in thermal equilibrium with a set of  nested  `heat reservoirs', where the  temperature of each reservoir is allowed to fluctuate owing to the interaction between adjacent  reservoirs  in the hierarchy.  By maximizing the entropy with respect to the conditional temperature distributions at each level of the hierarchy, subject to certain physically motivated constraints, we obtain a general family of distributions that includes two particular classes, namely  the {\it generalized gamma} and the {\it generalized inverse-gamma} distributions. 

The marginal distribution of  temperature of the innermost reservoir (i.e., at the lowest level of the hierarchy)  is  obtained by integrating  over the conditional distributions of all larger scales. Remarkably,  the resulting distribution  can be  written explicitly  in terms of a known special function, namely the Fox $H$-function. Averaging the quasi-equilibrium Boltzmann distribution of the small system over the temperature of the innermost reservoir  then yields the marginal distribution of states, which can also be written in terms of Fox $H$-functions. Here again the  distributions of states can be classified into two classes according to the tail behavior, namely the power-law and stretched-exponential classes. For a particular choice of constraints  our generalized distributions recover the distribution obtained in Ref.~\cite{ourPRE2017} in terms of Meijer $G$-functions. The H-theory described here  thus provides a rather general framework to describe  the statistics of fluctuations in complex systems with multiple time and length scales.

\section{Multiscale Systems}
\label{sec:ms}

We consider a multiscale complex system  that is  characterized by $N$ well-separated time scales, $\tau_i$, $i=1,...,N$, in addition to a 
large decorrelation time $\tau_0$  above which fluctuations in the system  are essentially uncorrelated.
 Let us order these timescales from smallest to largest:
$\tau_N\ll \tau_{N-1}\ll \cdots\ll \tau_1\ll \tau_0$.  Thus, if one samples the system at time intervals larger than or comparable to $\tau_0$, one will find the usual canonical distribution of states:
$p(\boldsymbol q|\beta_0)={\exp(-\beta_0 E(\boldsymbol q))}/{Z(\beta_0)}$, where $\boldsymbol q$ denotes the state variables, $\beta_0=1/k_BT_0$, with $T_0$ representing the `global' temperature of the system,  $E(\boldsymbol q)$ is the energy of the state labeled by $\boldsymbol q$, and $Z(\beta_0)$ is the large-scale partition function defined by $Z(\beta_0)=\int \exp(-\beta_0 E(\boldsymbol q)) d \boldsymbol q$.

At short time scales  (say, smaller than the smallest characteristic time $\tau_N$), the distribution of states  $p(\boldsymbol q)$ deviates considerably from the large-scale distribution $p(\boldsymbol q|\beta_0)$, owing to the complex  multiscale dynamics  of the system. In this scenario, it is  convenient  to consider the  system as being composed of a small subsystem---corresponding to the effective region where the measurements are performed---and  a large  subsystem  that has a slow internal dynamics characterized by several, hierarchically arranged timescales. Thus, in contrast to the usual canonical formulation, the large subsystem  can no longer  be treated  as a single `heat reservoir' with a fixed temperature. Instead,  it must  be viewed as a set of $N$  `nested  reservoirs' where each   reservoir is  described by a fluctuating temperature $T_j$, $j=1,...,N$. Physically, the fluctuations in these `local temperatures' are caused by the  interaction (exchange of energy) between  adjacent subsystems in the hierarchy, in analogy with the phenomenon of intermittency in turbulence \cite{frisch}.

Invoking Bayes's theorem, the joint equilibrium distribution $p(\boldsymbol q,\beta_1,...,\beta_N)$, where $\beta_j=1/k_BT_j$, can be factorized as
\begin{equation}\label{Njoint}
p(\boldsymbol q,\boldsymbol\beta)=p(\boldsymbol q|\boldsymbol\beta)p(\boldsymbol\beta).
\end{equation}
where we introduced the notation $\boldsymbol\beta\equiv(\beta_1,\beta_2,...,\beta_N)$.
Because of the hierarchical nature of our system, we assume that the conditional  distribution $p(\boldsymbol q|\boldsymbol\beta)$  depends only on the inverse temperature $\beta_N$  of the innermost reservoir, since this   is the only reservoir in `direct contact' with the  small subsystem, and so we write
\begin{align}
 p(\boldsymbol q|\boldsymbol\beta)=p(\boldsymbol q|\beta_N).
 \label{eq:pqb}
 \end{align}
 This means that the physical constraints imposed on the system  at the large scale (and which fix the global temperature $T_0$) are not directly felt at the small scales  but  rather are transferred down the hierarchy through the intervening  scales. 
 Under these assumptions, the marginal distribution $p(\boldsymbol q)$ can be written as
 \begin{align}
 P(\boldsymbol q)=\int_0^\infty P(\boldsymbol q|\beta_N)p(\beta_N)d\beta_N,
 \label{eq:1}
 \end{align}
where   the probability distribution $p(\beta_N)$ of the local inverse temperature $\beta_N$ is given by
\begin{align}
p(\beta_N) &= \int_{0}^{\infty}\cdots\int_{0}^{\infty}  p(\boldsymbol \beta) d\beta_1\cdots d\beta_{N-1}.
 \label{eq:pe1}
\end{align}

Owing to the separation of timescales, it is reasonable to assume that the small subsystem, which has a fast dynamics, is in local equilibrium with its immediate vicinity whose inverse temperature  $\beta_N$ changes much more slowly. In other words,   over short time periods (during which $\beta_N$ does not change appreciably)  the conditional probability $p(\boldsymbol q|\beta_N)$  can be described  by a Boltzmann distribution:
\begin{equation}\label{2MB}
p(\boldsymbol q|\beta_N)=\frac{\exp(-\beta_N E(\boldsymbol q))}{Z(\beta_N)}.
\end{equation}
The remaining task then  is  to find the  distribution $ p(\beta_N)$ of the local inverse temperature which encodes the complex dynamics of the multiscale background. 
This can be done by exploiting the hierarchical structure of the system, as argued below.

We assume  that a  subsystem (reservoir) at a given level $j$ of the hierarchy  interacts only with the reservoir at the next level up the hierarchy, and so we write the joint distribution $p(\boldsymbol\beta)$ as
\begin{equation}\label{3joint2}
p(\boldsymbol\beta)=
\prod_{j=1}^Nf(\beta_j|\beta_{j-1}),
\end{equation}
where $f(\beta_j|\beta_{j-1})$ denotes the probability density of  $\beta_j$ conditioned on a fixed value of  $\beta_{j-1}$.
In view of (\ref{eq:pe1}) and (\ref{3joint2}), the marginal distribution $p(\beta_N)$ can now be written as
\begin{align}
p(\beta_N) 
 &= \int_{0}^{\infty}\cdots\int_{0}^{\infty} \prod_{j=1}^{N}f(\beta_j|\beta_{j-1})d\beta_1\cdots d\beta_{N-1}, 
 \label{eq:pe}
\end{align}
In this way, our task has been reduced to computing the conditional distributions $f(\beta_j|\beta_{j-1})$, for $j=1,...,N$. In the next section we shall use a maximum entropy approach to solve this problem.

\section{Entropy Formulation}

\subsection{Multiscale entropy}

As usual, we define the information entropy of the joint distribution $p(\boldsymbol q,\boldsymbol\beta)$ by
\begin{align}\label{2entropyA}
S[p(\boldsymbol q,\boldsymbol\beta)]=-\int\int p(\boldsymbol q,\boldsymbol\beta)\ln p(\boldsymbol q,\boldsymbol\beta){d\boldsymbol q} d\boldsymbol\beta,
\end{align}
where we use the shorthand notation $d\boldsymbol \beta = \prod_{j=1}^Nd\beta_j$. In view of (\ref{Njoint}), (\ref{eq:pqb})  and (\ref{3joint2}), the entropy (\ref{2entropyA}) can be rewritten as
\begin{align}\label{2entropyB}
S[p(\boldsymbol q,\boldsymbol\beta)]
&=\int p(\boldsymbol\beta)s(\beta_N){d\boldsymbol\beta}-\sum_{k=1}^{N}\int p(\boldsymbol\beta)\ln f(\beta_k|\beta_{k-1})d \boldsymbol\beta_,
\end{align}
where $s(\beta_N)$ is the thermodynamic entropy of the small subsystem:
\begin{eqnarray}\label{2thermodentropy}
s(\beta_N)=
-\int p(\boldsymbol q|\beta_N)\ln p(\boldsymbol q|\beta_N)d\boldsymbol q,
\end{eqnarray}
which is a multiscale generalization of the entropy described in superstatistics \cite{beckpre2007,Dixit} for the case $N=1$. Let us also define the entropy at level $j$, for $j=0,...,N-1$, as the average of  $s(\beta_N)$ over all scales below this level, that is,
\begin{eqnarray}\label{sj}
s(\beta_j)=
\int s(\beta_N) p(\boldsymbol\beta)d \beta_{j+1}\cdots d\beta_N.
\end{eqnarray}
We  now seek to  maximize (\ref{2entropyB})  with respect to (w.r.t.) the distributions $f(\beta_j|\beta_{j-1})$. To this end, let us  first discuss  the constraints under which we shall carry out this maximization procedure. 

 \subsection{Constraints}


The first set of constraints is given by   the  usual normalization condition  
\begin{align}\label{Nnormalization}
\int  f(\beta_j|\beta_{j-1})d\beta_j=1, \qquad j=1,...,N.
\end{align}
The second set of constraints entails the choice of a moment to be kept fixed in the maximization procedure. Usually, the first moment (mean)  is the preferred choice \cite{VakarinPRE2006, Dixit}. Here, however, we shall adopt  a more general approach and    fix the $r$-th moment of the distributions $f(\beta_j|\beta_{j-1})$. More specifically, we require that
\begin{eqnarray}\label{3rmomentum}
\int\beta_j^r f(\beta_j|\beta_{j-1})d\beta_j=\beta_{j-1}^r,  \qquad j=1,...,N,
\end{eqnarray}
for some arbitrary  real $r\neq 0$ (not necessarily an integer).  Notice that (\ref{3rmomentum})  implies that
\begin{equation}\label{4rmomentum}
\langle \beta_j ^r\rangle\equiv \int\beta_j^r p(\boldsymbol\beta_j)d\boldsymbol\beta_j=\beta_0^r, \qquad j=1,...,N,
\end{equation}
where we introduced the notation
\[\boldsymbol\beta_j\equiv(\beta_1,...,\beta_j).\]
Eq.~(\ref{4rmomentum})  can be seen as a generalized equilibrium condition in the sense that the average value of $ \beta_j ^r$ is the same  at all levels of the hierarchy. 

As an additional constraint we  use the average entropy
\begin{eqnarray}\label{smomentum}
\langle s(\beta_N)\rangle \equiv \int\ s(\beta_N) p(\boldsymbol\beta)d\boldsymbol\beta = s(\beta_0),
\end{eqnarray}
where $s(\beta_0)$ is fixed. 
It then follows from  definition (\ref{sj}) that  the average entropy is the same across all scales:
\begin{eqnarray}\label{sjmomentum}
\langle s(\beta_j)\rangle \equiv \int\ s(\beta_j) p(\boldsymbol\beta_j)d\boldsymbol\beta_j = s(\beta_0),  \qquad j=1,...,N,
\end{eqnarray}
which is a reasonable  equilibrium condition. Furthermore, we shall assume that the thermodynamic entropy defined in (\ref{2thermodentropy}) satisfies the following relation
\begin{align}
s(\beta_N) \sim s_0 \ln \beta_N,
\label{eq:sN}
\end{align}
where $s_0$ is a constant and the notation $\sim$ indicates equality except for an  additive constant. (In other words,  $f(x)\sim g(x)$ means here that $f(x)=g(x)+C$, where $C$ is a constant.) We recall that relation (\ref{eq:sN}) is valid for a large class of systems, such as  those that obey the equipartition theorem, for which the internal energy is proportional to the temperature  \cite{VakarinPRE2006,Dixit}. 

We also make the  assumption that  the distribution $f_{k}(\beta_k|\beta_{k-1})$ 
 is  invariant under a rescaling of the variables $\beta \rightarrow \lambda \beta$:
\begin{equation}\label{scale}
\begin{split}
f_{k}(\beta_k|\beta_{k-1})d\beta_k=f_k(\lambda\beta_k|\lambda\beta_{k-1})d(\lambda\beta_k) .
\end{split}
\end{equation}
Physically, this  means that  the temperature distributions should remain of the same form regardless of the   temperature scale one chooses.
Now, if we make $\lambda=1/\beta_{k-1}$  in (\ref{scale}) we  get
\begin{equation}\label{scale2}
\begin{split}
f_{k}(\beta_k|\beta_{k-1})d \beta_k=g_k\left(\frac{\beta_k}{\beta_{k-1}}\right)\frac{d\beta_k}{\beta_{k-1}}= g_k(u) du,
\end{split}
\end{equation}
for some function $g_k(u)$, where $u=\beta_k/\beta_{k-1}$. 
Relation  (\ref{scale2}) leads to the following two useful relations that are proven in Appendix A:
\begin{align}\label{eq:Sn1}
\int p(\boldsymbol\beta_k) \ln  \beta_k d{\boldsymbol\beta_k}\; \sim  \int p(\boldsymbol\beta_j) \ln \beta_{j}  d\boldsymbol\beta_j,\qquad \mbox{for $j\le k$},
\end{align} 
and
\begin{align}\label{eq:Sn3}
&\int p(\boldsymbol\beta_k)\ln f(\beta_{k}|\beta_{k-1}) d{\boldsymbol\beta_k}\; \sim  - \int p(\boldsymbol\beta_j) \ln \beta_{j}  d\boldsymbol\beta_j,\qquad \mbox{for $j<k$}.
\end{align}

Now, inserting (\ref{eq:sN}) into (\ref{sj}) and using (\ref{eq:Sn1}), one  finds that
\begin{align}
s(\beta_j) =s_0 \ln \beta_j +s_j,
\label{eq:sjln}
\end{align}
where $s_j$ is a constant that does not depend on $\beta_j$. In view of this relation, the constraint (\ref{sjmomentum}) can be  written as
\begin{align}\label{logbetaj}
\int \left(\ln \beta_j \right) p(\boldsymbol\beta_j)d\boldsymbol\beta_j=c_j,
\end{align}
where $c_j$ is a constant.

\subsection{Entropy maximization}
\label{sec:max}

In order to maximize (\ref{2entropyB}) w.r.t.~$f(\beta_j|\beta_{j-1})$, for any given $j$, it is necessary to make it explicit the dependence of $S[p(\boldsymbol q,\boldsymbol\beta)]$ on $f(\beta_j|\beta_{j-1})$. To this end, 
we  first note that on use of (\ref{3joint2}) and (\ref{Nnormalization}) we can rewrite (\ref{2entropyB})  as
\begin{align}\label{2entropyC}
S[p(\boldsymbol q,\boldsymbol\beta)]
&= \int s(\beta_j) p(\boldsymbol\beta_j){d\boldsymbol\beta_j} -
\sum_{k=1}^{j-1}\int p(\boldsymbol\beta_k)\ln f(\beta_{k}|\beta_{k-1})d{\boldsymbol\beta_k}
 \cr & - \int p(\boldsymbol\beta_j)\ln f(\beta_{j}|\beta_{j-1})d{\boldsymbol\beta_j}- \sum_{k=j+1}^{N}\int p(\boldsymbol\beta_k)\ln f(\beta_{k}|\beta_{k-1})d{\boldsymbol\beta_k}.
\end{align}
Now using  (\ref{eq:Sn3}) and  (\ref{eq:sjln})  in (\ref{2entropyC}), one  finds that
\begin{align}\label{4entropy}
S[p(\boldsymbol q,\boldsymbol\beta)]&\sim c_j \int p(\boldsymbol\beta_j)\ln \beta_{j}{d\boldsymbol\beta_j}
-\int p(\boldsymbol\beta_j) \ln f(\beta_{j}|\beta_{j-1})d\boldsymbol\beta_j -\sum_{k=1}^{j-1}\int p(\boldsymbol\beta_k)\ln f(\beta_k|\beta_{k-1})d\boldsymbol\beta_k,
\end{align}
where $c_j=N-j-s_0$.
 Note that the entropy $S[p(\boldsymbol q,\boldsymbol\beta)]$ depends on $f(\beta_j|\beta_{j-1})$ only through the first two terms in the right hand side of (\ref{4entropy}). 

Maximizing (\ref{4entropy}) w.r.t.~$f(\beta_j|\beta_{j-1})$, subject to the constraints (\ref{Nnormalization}), (\ref{3rmomentum}) and (\ref{logbetaj}), yields
\begin{equation}
\int \left[\ln f(\beta_j|\beta_{j-1}) +A_j + B_j\beta_j^r +C_j  \ln\beta_j\right]\delta_j p(\boldsymbol\beta_j)d\boldsymbol\beta_j=0,
\label{eq:maxN}
\end{equation}
where $A_j$, $B_j$, and $C_j$  are Lagrange multipliers and $\delta_j p(\boldsymbol\beta_j)\equiv p(\boldsymbol\beta_{j-1})\delta f(\beta_j|\beta_{j-1})$. 
The solution to  (\ref{eq:maxN})  takes the form
\begin{equation}\label{tempgengamma}
f(\beta_j|\beta_{j-1})=e^{-A_j}\beta_{j}^{-C_j}\exp{\left(-B_j\beta_j^r\right)}.
\end{equation}
To enforce the constraint
 (\ref{3rmomentum}) we choose  $B_j=\alpha_j/\beta_{j-1}^r$ and set $C_j=-r\alpha_j+1$, where $\alpha_j>0$.
 Using  these parameters in (\ref{tempgengamma}) one obtains the following general distribution:
\begin{equation}\label{gengammaj}
f_{j}(\beta_j|\beta_{j-1})=\frac{|r|\alpha_{j}^{\alpha_j}}{\beta_j\Gamma(\alpha_j)}\left(\frac{\beta_j}{\beta_{j-1}} \right)^{r\alpha_j}\exp\left[{-\alpha_j\left(\frac{\beta_{j}}{\beta_{j-1}}\right)^r}\right].
\end{equation}
For $r>0$ this distribution corresponds to  the {\it generalized gamma distribution}, whereas  for $r<0$ it gives the {\it generalized inverse gamma distribution}.

We note furthermore that for the particular case $r=1$ the distribution (\ref{gengammaj}) yields the usual gamma distribution,
\begin{equation}\label{gamma}
f_{j}(\beta_j|\beta_{j-1})=\frac{\left(\alpha_{j}/\beta_{j-1}\right)^{\alpha_j}}{\Gamma(\alpha_j)}{\beta_j}{} ^{\alpha_j-1}\exp\left({-\frac{\alpha_j\beta_{j}}{\beta_{j-1}}}\right),
\end{equation}
whereas for $r=-1$  it gives the standard  inverse gamma distribution:
\begin{equation}\label{invgamma}
f_{j}(\beta_j|\beta_{j-1})=\frac{\left(\alpha_{j}\beta_{j-1}\right)^{\alpha_j}}{\Gamma(\alpha_j)}{\beta_j}^{-\alpha_j-1}\exp\left({-\frac{\alpha_j\beta_{j-1}}{\beta_{j}}}\right).
\end{equation}

It is interesting to note that the generalized inverse gamma distribution has recently been used to model  wealth distribution in  ancient Egypt \cite{egypt}. The  Weibull and the Frechet distributions, which are particular cases of the generalized gamma and generalized inverse-gamma distributions, respectively, have also found important applications in extreme value statistics \cite{extreme_val} and sum of correlated random variables \cite{Weibull}. Here, however, our interest is  to use  (\ref{gengammaj})   not so much as a standalone distribution but rather as a means to obtain  the   distribution $p(\beta_N)$ of inverse temperatures at the  innermost reservoir, from which the  distribution of states $p(\boldsymbol q$)  can   be found. This is done next.

\section{The equilibrium distributions}

As discussed in Sec.~\ref{sec:ms}, the complex  dynamics of the  large system (background) is felt by the small subsystem only through the fluctuations of the  inverse temperature $\beta_N$  of the innermost reservoir. Thus, in order to determine the marginal distribution of states $p(\boldsymbol q)$ of the small subsystem, it  is   necessary first to compute the distribution $p(\beta_N)$; see (\ref{eq:1}). It is remarkable that  both these distributions can be obtained in analytical form in terms of some special transcendental functions known as the Fox $H$-functions \cite{Hfunction}, as shown below.

\subsection{The background distribution}

The marginal distribution $p(\beta_N)$ at the lowest level of the hierarchy  is given by   (\ref{eq:pe}), where each of the distributions $f(\beta_{j}|\beta_{j-1})$ appearing in this   expression   is as shown  in (\ref{gengammaj}).  In computing  the multiple integrals in (\ref{eq:pe})  the cases  $r>0$ and $r<0$ need to be treated separately, but for both  cases these  integrals  can be calculated explicitly in terms of  the Fox $H$-functions. 

As shown in Appendix B, for the  case $r>0$ one finds
\begin{equation}
\label{H-gamma}
    p(\beta_N )=
\omega_\rho\Omega    H_{ 0,N } ^{ N,0 }  \left( 
\begin{array}{c}
{-} \\ 
{ (\boldsymbol\alpha-\rho {\bf 1},\rho {\bf 1})}
\end{array}
\bigg |\frac{\omega_\rho \beta_N}{\beta_0 }  \right),
\end{equation}
whilst for $r<0$ the result is
\begin{equation}
\label{H-inv-gamma}
    p(\beta_N )=
\frac{\Omega}{\omega_\rho}    H_{ N,0 } ^{ 0,N }  \left( 
\begin{array}{c}
{ ((1-\rho) {\bf 1}-\boldsymbol\alpha,\rho{\bf 1})} \\ 
{-}
\end{array}
\bigg |\frac{\beta_N}{\omega_\rho\beta_0 }  \right),
\end{equation}
where $\rho=1/|r|$, $\omega_\rho  =\prod_{j=1}^{N}\alpha_j^{\rho}$, and $\Omega=1/\left(\beta_0\Gamma(\boldsymbol\alpha)\right)$. Here we have introduced the vector notation
${\boldsymbol\alpha}\equiv (\alpha_1,\dots,\alpha_N)$ and 
$
\Gamma({\bf a})  \equiv\prod_{j=1}^{N}\Gamma (a _j)$.  We have also used a dash   in the top row of the  $H$-function in (\ref{H-gamma}) and in the low row of  the $H$-function in (\ref{H-inv-gamma})  to indicate  that the respective parameters are not present.

We note in passing that after setting $|r|=1$  in  expressions (\ref{H-gamma}) and  (\ref{H-inv-gamma}) we recover the two classes of universality for the background distributions obtained in Ref.~\cite{ourPRE2017} from a stochastic dynamical model. To see this, we  note that for  $\rho=1$ the set of parameters  $\rho{\bf 1}\equiv(\rho,...,\rho)$ appearing in each of the $H$-functions above becomes simply the identity vector, in which case the $H$-function  reduces to a simpler function, namely the Meijer $G$-function \cite{Hfunction}.
Setting   $\rho=1$ in   (\ref{H-gamma})  then yields  
\begin{equation}
\label{meijer2}
     p(\beta_N )=
\omega\Omega   G_{ 0,N } ^{ N,0 }  \left( 
\begin{array}{c}
{-} \\ 
{ \boldsymbol\alpha-{\bf 1}}
\end{array}
\bigg |\frac{\omega \beta_N}{\beta_0 }  \right),
\end{equation}
whilst  from (\ref{H-inv-gamma})  one has
\begin{equation}
\label{meijer1}
     p(\beta_N )= \frac{\Omega}{\omega}    G_{ N,0 } ^{ 0,N }  \left( 
\begin{array}{c}
{- \boldsymbol\alpha}\\ 
{-}
\end{array}
\bigg |\frac{ \beta_N}{\beta_0 \omega} \right),
\end{equation}
where $\omega   =\prod_{j=1}^{N}\alpha_j$. In  comparing  the distributions (\ref{meijer2})  and (\ref{meijer1}) with the corresponding expressions given Ref.~\cite{ourPRE2017} 
one has to bear in mind that there the distributions are written in terms of a variable $\varepsilon_N$ which corresponds in the notation of the present paper to $1/\beta_N$.

\subsection{The  distribution of states}

In view of  (\ref{eq:1}) and (\ref{2MB}),  the marginal distribution of states  $p(\boldsymbol q)$
 of  the small subsystem  can be written as
\begin{align}\label{eq:px}
p(\boldsymbol q)&=\int_0^\infty \frac{\exp(-\beta_N E(\boldsymbol q))}{Z(\beta_N)}p(\beta_N)d \beta_N
\end{align}
where $p(\beta_N)$ is given by either (\ref{H-gamma}) or (\ref{H-inv-gamma}). In order to carry out this integral one needs to know the dependence of the partition function $Z(\beta_N)$ on $\beta_N$. In view of the fact that $S(\beta)\sim \ln Z(\beta)$, it then follows from assumption (\ref{eq:sN})  that $Z(\beta_N) \sim\beta_N^{-\gamma}$, for some exponent $\gamma>0$, and so we write
\begin{align}\label{eq:zn}
Z(\beta_N) = Z(\beta_0) \left(\frac{\beta_N}{\beta_0}\right)^{-\gamma}.
\end{align}
Inserting (\ref{eq:zn}) into (\ref{eq:px}) yields
\begin{align}\label{eq:px2}
p(\boldsymbol q)&=\frac{1}{Z(\beta_0)} \int_0^\infty \left(\frac{\beta_N}{\beta_0}\right)^{\gamma} \exp\left(-\beta_N E(\boldsymbol q)\right) p(\beta_N)d \beta_N
\end{align}
It is also remarkable that this integral can be carried out explicitly in terms of Fox $H$-functions for both classes of  background distributions, with the resulting distributions   being classified  into two classes according to the behavior at the tails, as follows:

\bigskip

 \noindent i) {\it Power-law class}. This is the case when $r>0$. Upon inserting (\ref{H-gamma}) into (\ref{eq:px2})  and using a convolution property of the $H$-function \cite{Hfunction}, the resulting integral can be performed explicitly (see Appendix C), yielding
\begin{align}
\label{HgammaX}
    p(E) =\frac{1}{Z(\beta_0)\omega_\rho^\gamma \Gamma(\boldsymbol\alpha)}    H_{ N,1 } ^{ 1,N }  \left( 
\begin{array}{c}
{((1-\gamma\rho){\bf 1}-\boldsymbol\alpha,\rho {\bf 1})} \\ 
{(0,1 )}
\end{array}
\bigg |\frac{\beta_0 E}{\omega_\rho }  \right).
\end{align}
Here we have omitted the state variable $\boldsymbol q$ for simplicity of notation, with the understanding that $p(E)$ denotes the probability of a state $\boldsymbol q$ with energy $E(\boldsymbol q)$. From the asymptotic expansion of the $H$-function for large arguments one finds \citep{Hfunction} that the $p(E)$ decays as a power-law for large values of $E$:
 \be 
p(E)\sim 
 \sum_{j=1}^{N} \frac{c_{j}}{E^{\gamma+|r|\alpha_j}}, \quad  \mbox{for}\quad E\to\infty,
 \label{eq:asym1}
\ee 
 where the $c_i$'s are constants. 
To illustrate the power-law class of distributions we show in Fig.~1 some plots of the function $p(E)$ given in (\ref{HgammaX}) for  cases where $\gamma=1$, $\beta_0=1$, $Z(\beta_0)=1$, and $\alpha_j=\alpha=1.0$. The  values of the parameters $N$ and $r$ for each plot is indicated in the  caption of the figure. The main plots in Fig.~1 are in semilogarithmic scale, whilst the insets show the same data in log-log scale. One clearly sees from Figs.~1(a) and 1(b) that the smaller the value of the parameter $r$, for $N$ fixed, the heavier the tail of the distribution. This is in agreement with the aympotic behavior given in (\ref{eq:asym1}) which  shows that the exponent of the power law decreases as $r$ decreases. Similarly, from Figs.~1(c) and 1(d) one sees that the larger the number $N$ of scales, for $r$ fixed, the heavier the tails. Note, however, that the exponent of the power-law  does not depend on $N$; see (\ref{eq:asym1}). It is instead the prefactor  that increases with $N$, since we are taking $\alpha_j=\alpha$, for $j=1,...,N$,  thus causing a slower decay of the tail.

\begin{figure}[t]
\label{fig:2}
\center \subfigure[\label{fig2a}]{\includegraphics[width=0.45\textwidth]{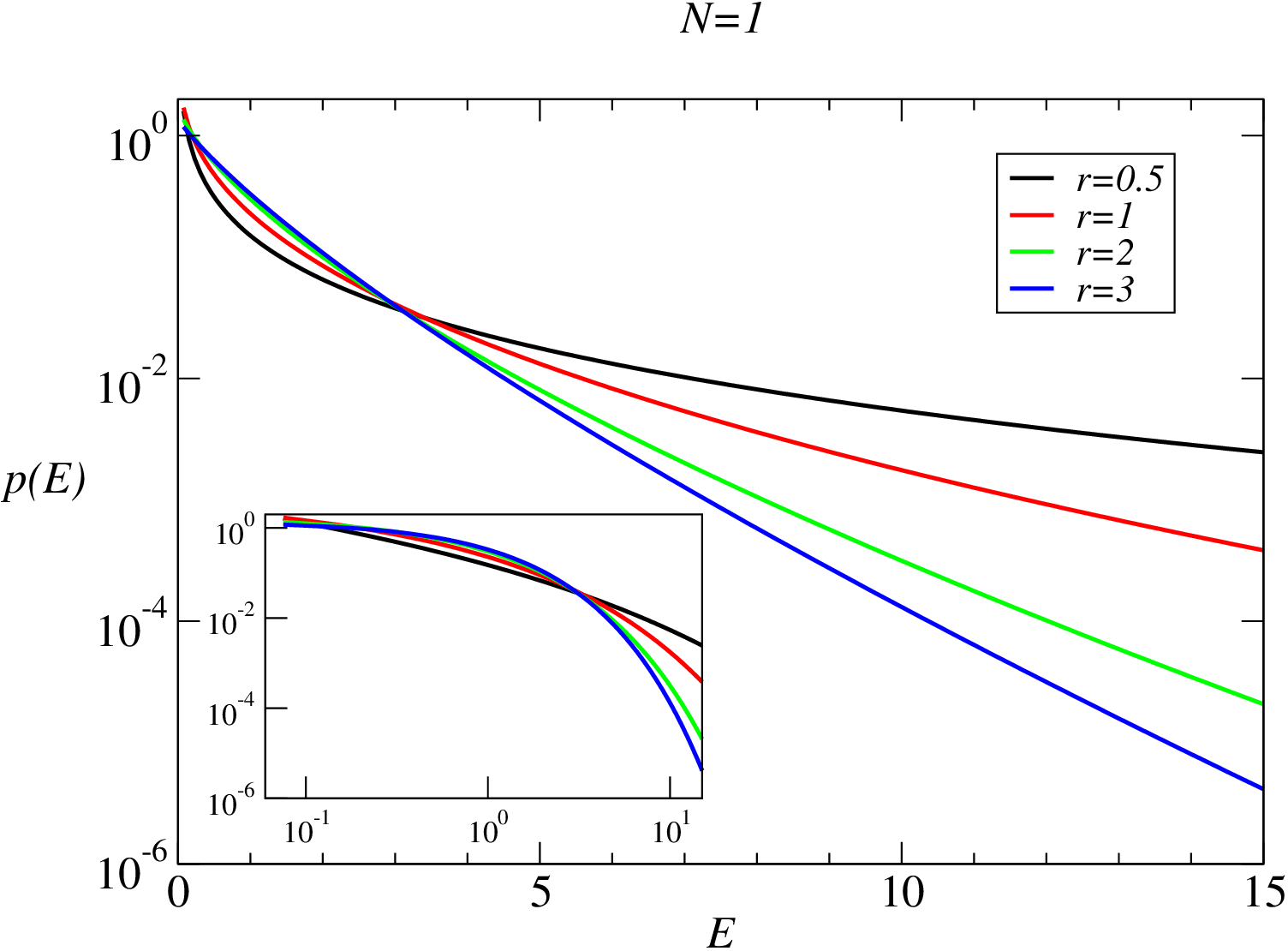}}
\quad \subfigure[\label{fig2b}]{\includegraphics[width=0.45\textwidth]{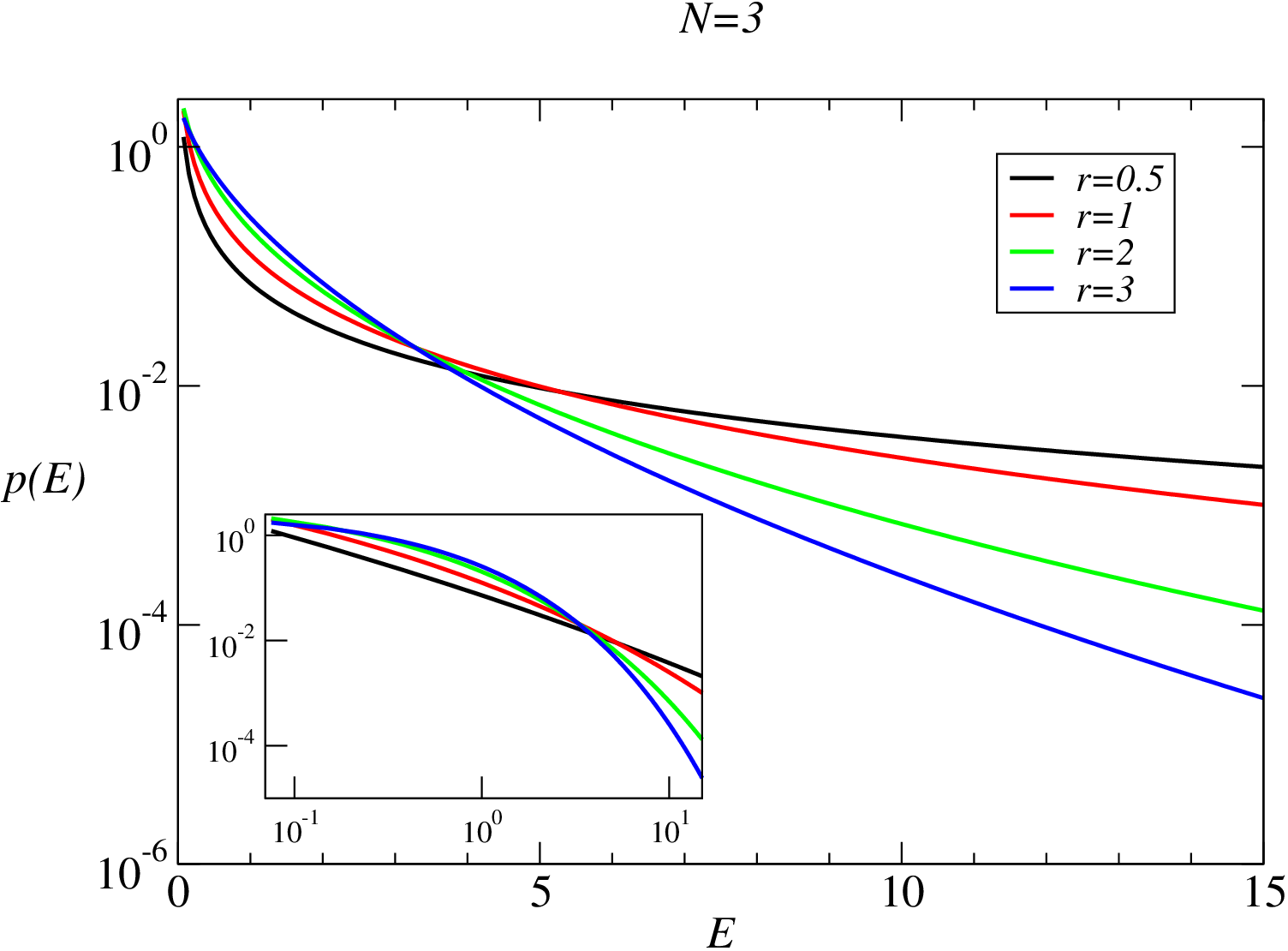}}
\center \subfigure[\label{fig2c}]{\includegraphics[width=0.45\textwidth]{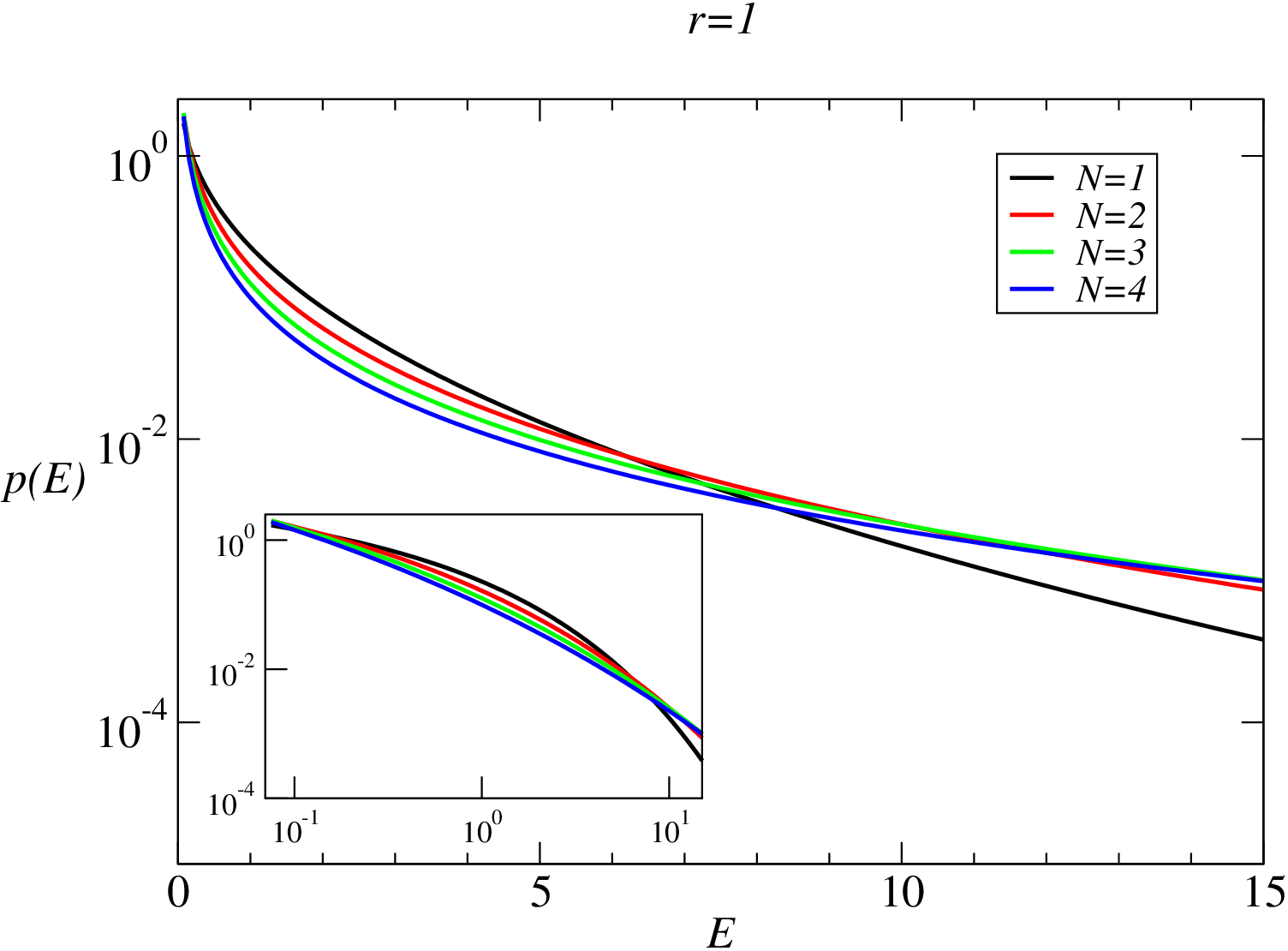}}
\quad \subfigure[\label{fig2d}]{\includegraphics[width=0.45\textwidth]{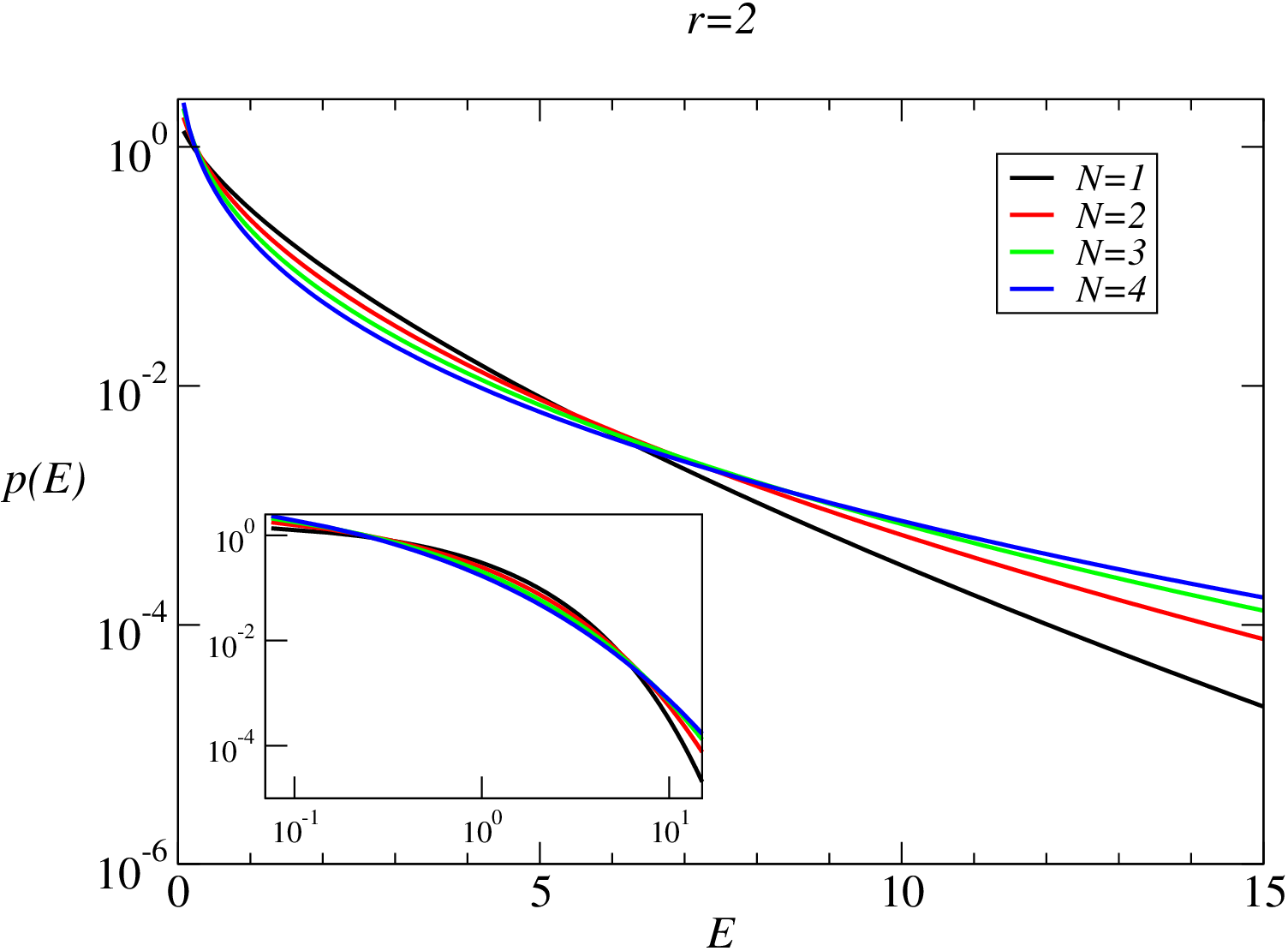}}
\caption{(Color online) (a)  Distribution of states $p(E)$ for the power-law class for the following values of parameters: (a) $N=1$, $r=0.5,1 , 2, 3$; (b) $N=3$, $r=0.5,1, 2, 3 $; (c) $r=1$, $N=1, 2, 3, 4$; and (d) $r=2$, $N=1, 2, 3, 4$. In all cases shown here we have used $\gamma=1$, $\beta_0=1$, $Z(\beta_0)=1$, and $\alpha_j=\alpha=1.0$, for $j=1,...,N$.}
\end{figure}

\begin{figure}[t]
\center \subfigure[\label{fig1a}]{\includegraphics[width=0.45\textwidth]{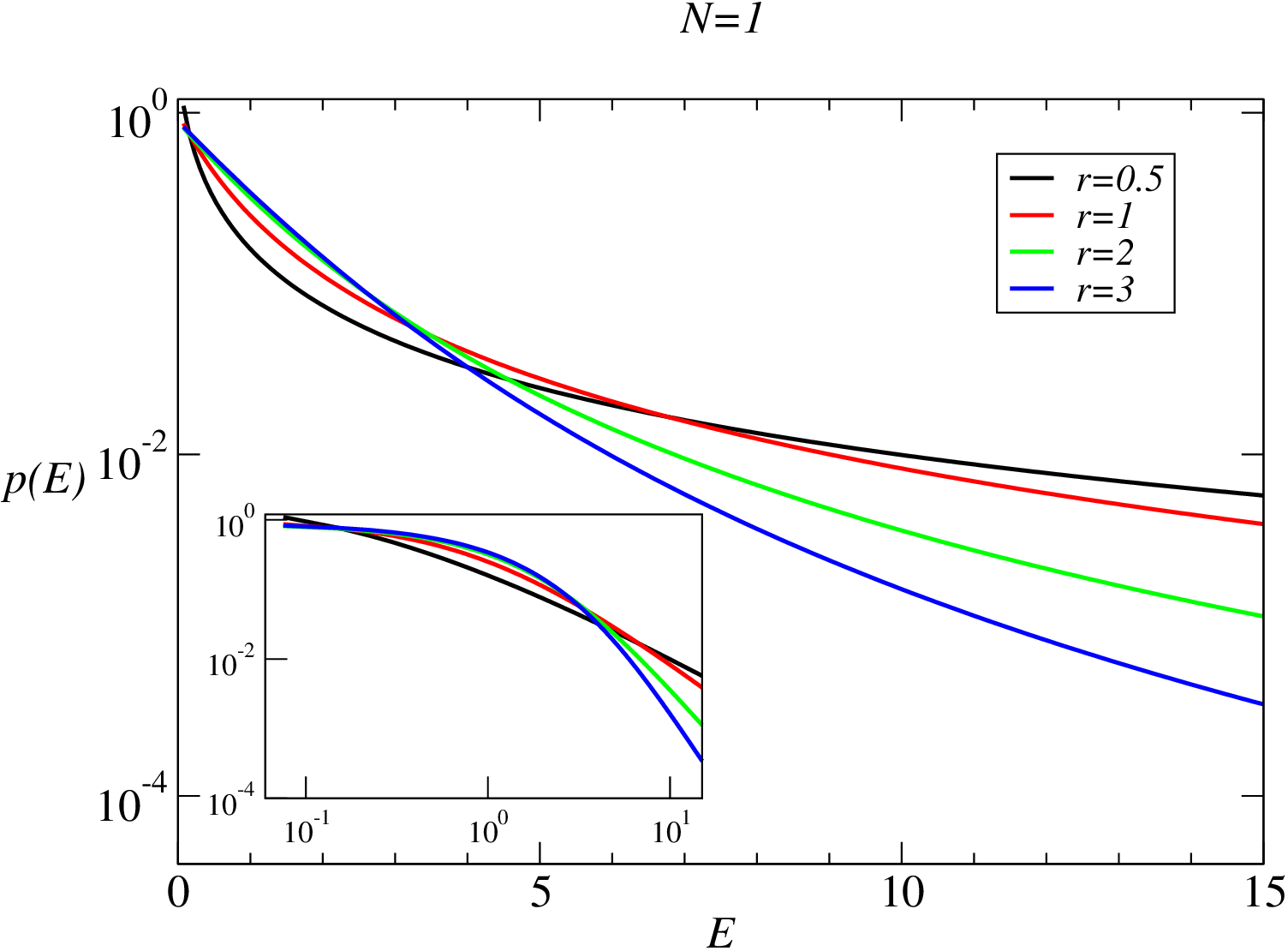}}
\qquad \subfigure[\label{fig1b}]{\includegraphics[width=0.45\textwidth]{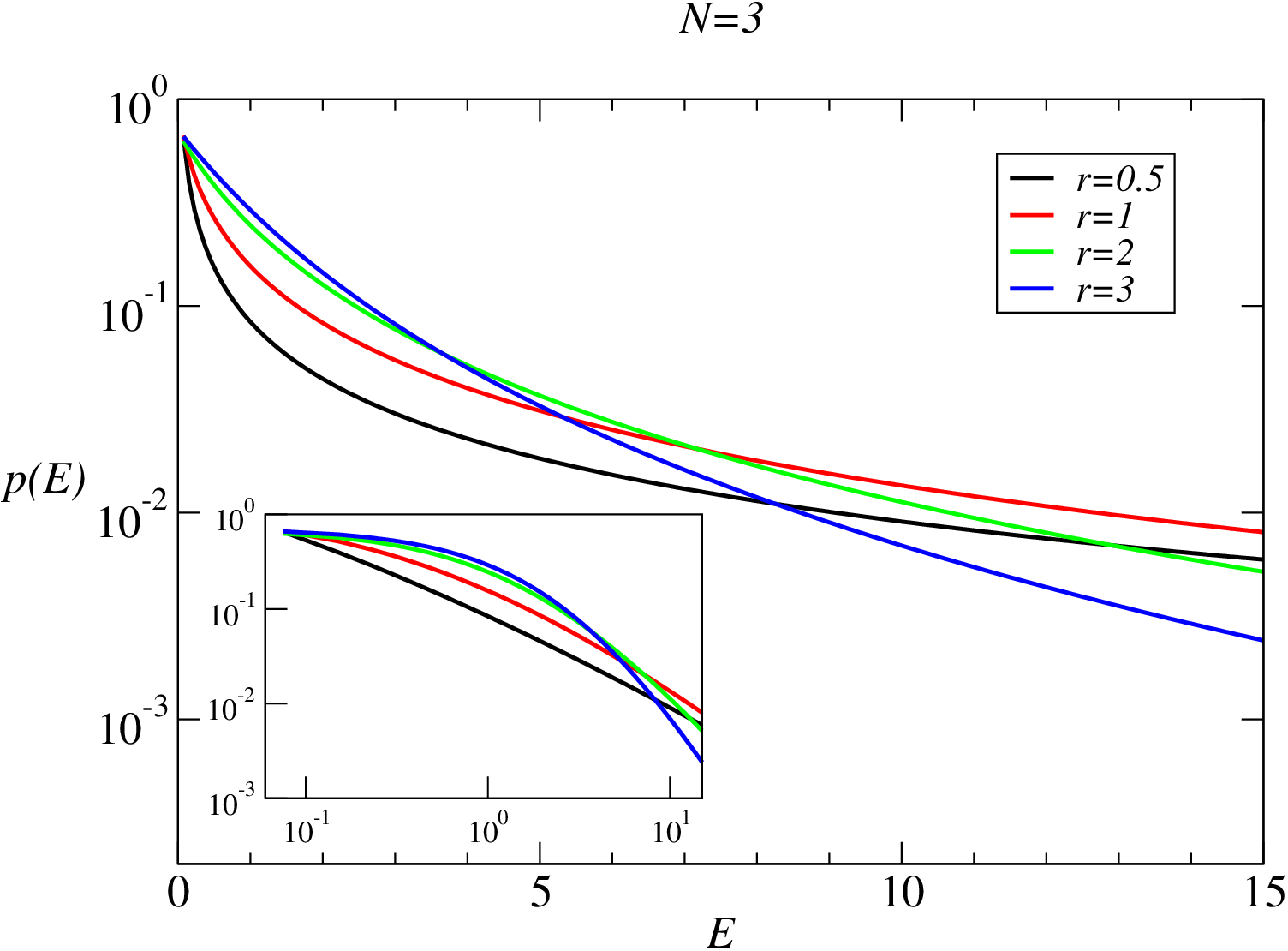}}
\center  \subfigure[\label{fig1c}]{\includegraphics[width=0.45\textwidth]{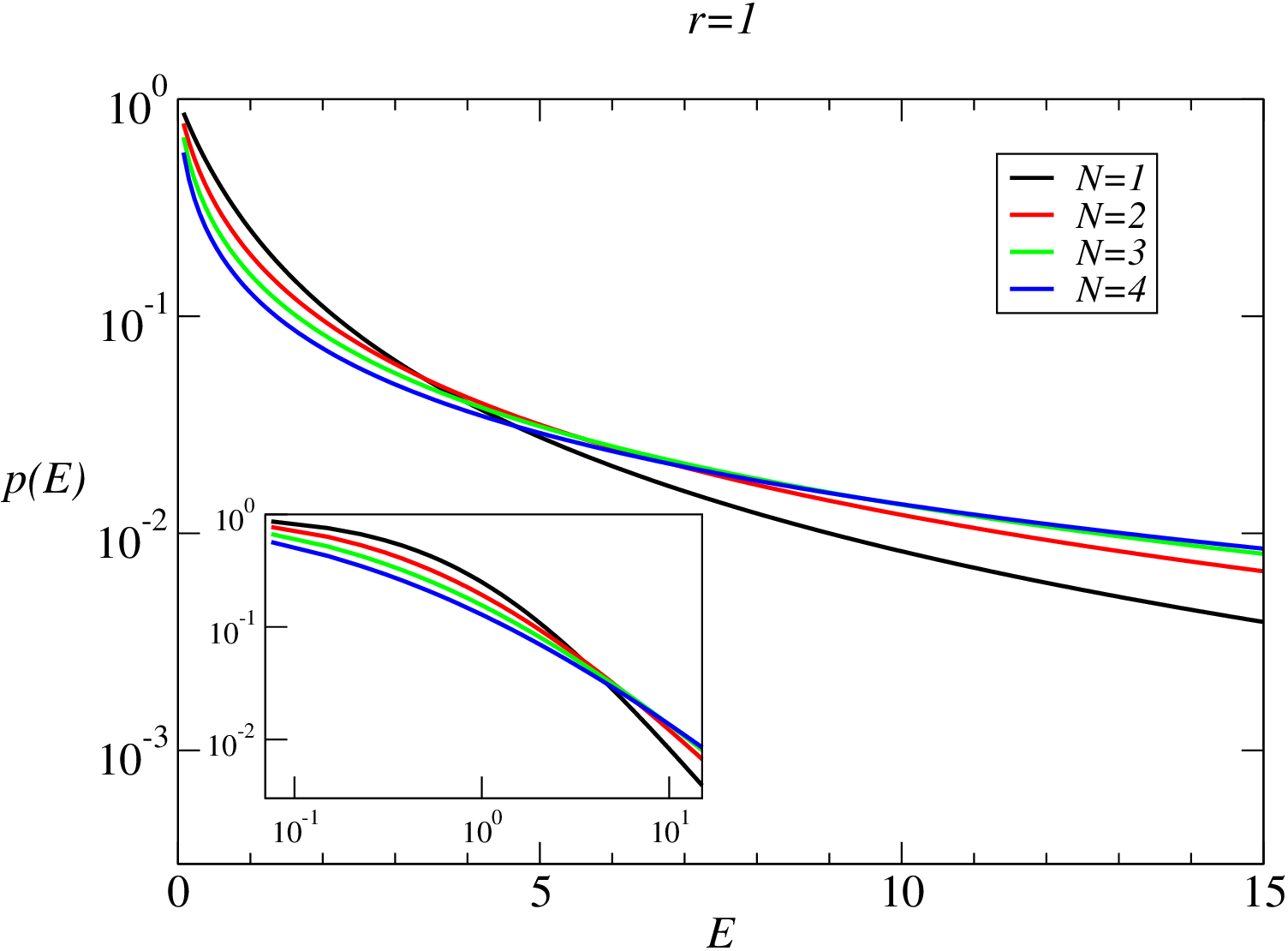}}
\qquad \subfigure[\label{fig1d}]{\includegraphics[width=0.45\textwidth]{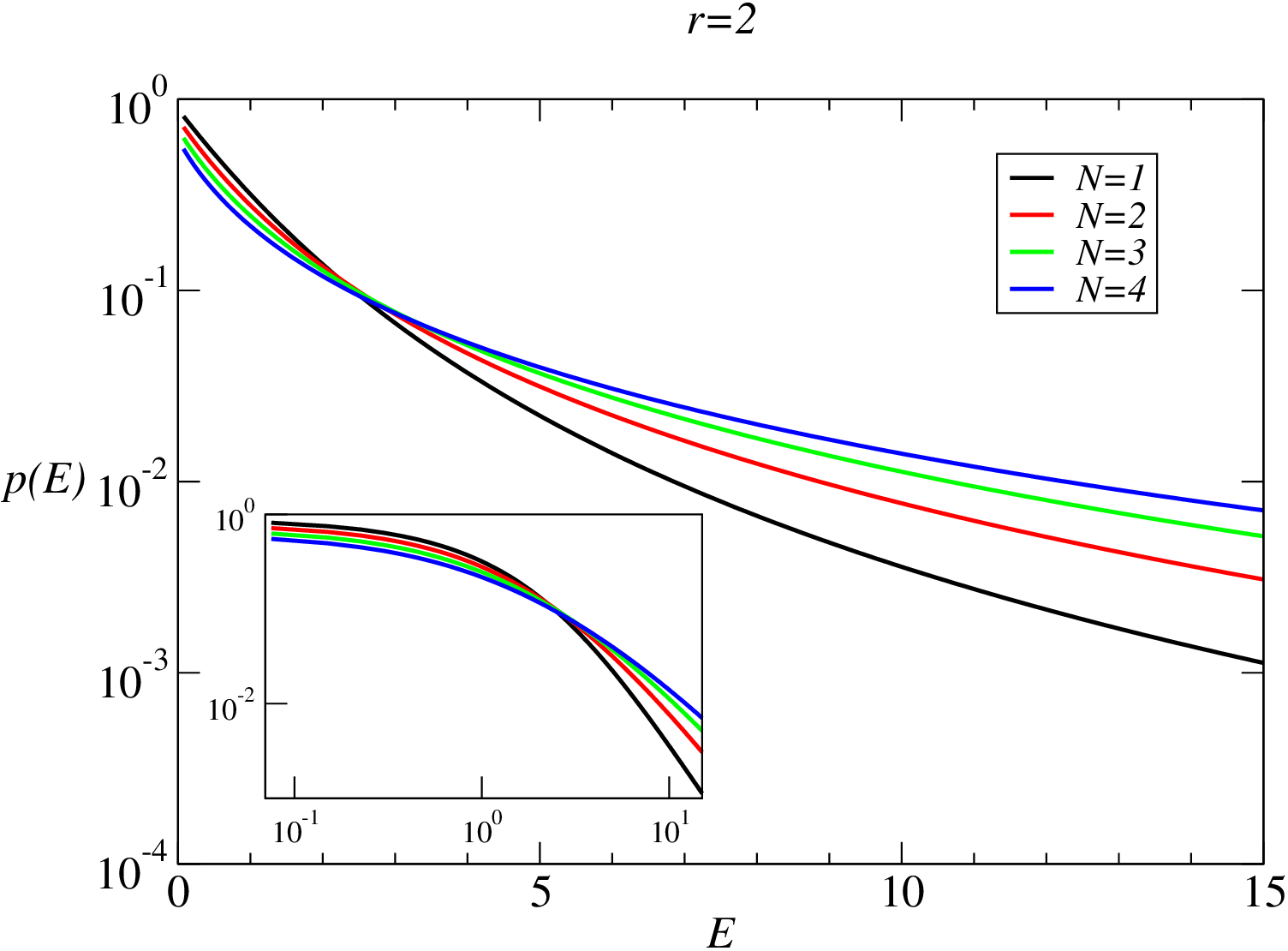}}
 \label{fig:1}
\caption{(Color online) Distribution of states $p(E)$ for the stretched exponential class with the same choice of parameters as in Fig.~1.}
\end{figure}

 \noindent ii) {\it Stretched-exponential class}. This corresponds to the case  $r<0$.
Here the integral (\ref{eq:px2}), with $p(\beta_N)$ as given  in (\ref{H-inv-gamma}),  can be written as
\begin{align}
\label{Hinv_gammaX}
    p(E)=
\frac{\omega_\rho^\gamma}{ Z(\beta_0)\Gamma(\boldsymbol\alpha)}    H_{ 0,N+1 } ^{ N+1,0 }  \left( 
\begin{array}{c}
{-}\\
{ (\boldsymbol\alpha-\gamma\rho {\bf 1},\rho{\bf 1}),(0,1)} 
\end{array}
\bigg |\omega_\rho\beta_0 E  \right),
\end{align}
as also shown in Appendix C. The asymptotic behavior in this case is given by a modified stretched exponential:
  \be \
p(E)\sim 
{E^{\theta}}{\exp\left[-A(\omega_\rho \beta_0 E)^{1/(\rho N+1)}\right]},
\quad  \mbox{for}\quad E \to\infty,
\label{eq:asym2}
\ee 
where $\theta=N(\bar{\alpha}-\gamma\rho-1/2)/(\rho N+1)$, $\bar{\alpha}=(1/N)\sum_{i=1}^N \alpha_i$ and $A=(\rho N+1)\rho^{-\rho N/(\rho N +1)}$. 
Some illustrative plots of the function $p(E)$ given in  (\ref{Hinv_gammaX}) are shown in  Fig.~2 for the same choice of parameters as in Fig.~1. The same qualitative dependence of the tails on the parameters $N$ and $r$ are observed here: the larger the value of $N$ or the smaller the choice of $r$, the heavier the tails. This behavior is in agreement with (\ref{eq:asym2}) which shows that the exponent of the stretched exponential decreases with both the  increase of $N$ and  the decrease of $r$.

We note in passing that the particular cases $r=\pm1$   yield results consistent with those obtained in Ref.~\cite{ourPRE2017}, in  that the corresponding  distributions can also be written in terms of $G$-functions. For $\rho=1$ the expression (\ref{HgammaX}) simplifies to 
\be 
 p(E)=
\frac{1}{Z(\beta_0)\omega^\gamma \Gamma(\boldsymbol\alpha)}G_{N,1}^{1,N}\left( 
\begin{array}{c}
{(1-\gamma){\bf 1}-\boldsymbol\alpha} \\ 
0
\end{array}
\bigg |\frac{\beta_0 E}{\omega}\right),
\label{eq:PN1}
\ee 
whereas the distribution (\ref{Hinv_gammaX}) reads
\be 
 p(E)=
\frac{\omega^\gamma}{ Z(\beta_0)\Gamma(\boldsymbol\alpha)}G_{0,N+1}^{N+1,0}\left( 
\begin{array}{c}
- \\ 
{ \boldsymbol\alpha-\gamma{\bf 1},0}
\end{array}
\bigg |\omega\beta_0 E\right) .
\label{eq:PN2}
\ee 
\par
 
\section{Conclusions}

In this paper, we have used  a maximum entropy principle to derive a generalized version of the multicanonical formalism (H-theory) introduced in Refs.~\cite{SV2,ourPRE2017}. 
In our approach the system is considered to be effectively composed of a small subsystem in thermal equilibrium with a hierarchical set of heat reservoirs, whose local temperatures fluctuate owing to weak interactions between adjacent reservoirs. We characterized the joint equilibrium distribution of the state variables and the local inverse temperatures  by means of its Shannon information entropy.  This entropy was  maximized with respect to the conditional temperature distributions at each level of the hierarchy, subject to certain physically motivated constraints.  The large family of distributions  that were found by this procedure can be grouped into two classes:  the {\it generalized gamma} and the {\it generalized inverse-gamma} distributions. The knowledge of these conditional distributions of inverse temperatures  allowed us to obtain the marginal distribution $p(\beta_N)$ of the  inverse temperature at the lowest level of the hierarchy, which was explicitly  written for both classes in terms of the Fox $H$-functions. 

The marginal distribution of states $p(\boldsymbol q)$ was then obtained by averaging the  conditional distribution of states $p(\boldsymbol q|\beta_N)$ over the local inverse-temperature $\beta_N$ and the resulting distribution was also  written in terms of Fox $H$-functions. These distributions exhibit heavy tails that can be classified into two classes, namely the power-law and stretched-exponential classes.  The distributions derived in Ref.~\cite{ourPRE2017} from a stochastic dynamical approach, which were written in terms of Meijer $G$-functions, were shown to be particular cases of the Fox $H$-functions obtained from the maximum entropy approach. The H-theory presented here  thus provides a rather general framework to describe  the statistics of fluctuations in complex systems with multiple time/space scales, quite irrespective of the detailed underlying dynamics. Applications of H-theory in the context of Eulerian and Lagrangian turbulence, mathematical finance and random lasers have had  great success. Further applications of the generalized formalism presented here to other complex systems with multiple spatio-temporal scales are under current investigation.

\begin{acknowledgments}
 This work was supported in part by the Conselho Nacional de Desenvolvimento Cient\'ifico e Tecnol\'ogico (CNPq), under Grants No.~308290/2014-3 and No.~311497/2015-2, and by FACEPE, under Grant No.~APQ-0073-1.05/15.  We thank W.~Sosa for generating the data used in the figures. 
\end{acknowledgments}

\appendix
\section{Derivation of (\ref{eq:Sn1}) and  (\ref{eq:Sn3})}

First consider a term of the form
\begin{align}
\int p(\boldsymbol\beta_{k}) \ln \beta_k\;  d \boldsymbol \beta_k.
\end{align}
This can  be rewritten as
\begin{align}
\int  p(\boldsymbol\beta_k)\ln  \beta_k \; d\boldsymbol\beta_k&= \int p(\boldsymbol\beta_{k}) \left[ \ln \left(\frac{\beta_k}{\beta_{k-1}}\right)+\ln  \beta_{k-1} \right]d\boldsymbol\beta_{k} .
\end{align}
Upon using property (\ref{scale2}) we then obtain
\begin{align}
\int  p(\boldsymbol\beta_{k})\ln  \beta_k d\boldsymbol\beta_{k}&=\left(\int   g_k(u) \ln u \, du\right) \int   p(\boldsymbol\beta_{k-1})  d\boldsymbol\beta_{k-1}+
 \int p(\boldsymbol\beta_{k-1}) \ln \beta_{k-1}  \; d\boldsymbol\beta_{k-1}
\cr
&=A_k  + \int  p(\boldsymbol\beta_{k-1}) \ln  \beta_{k-1} \; d\boldsymbol\beta_{k-1}
\label{eq:Sn7}
\end{align}
where  $A_k=\int_0^\infty   g_k(u) \ln u  \, du$ is a constant. This implies that 
\begin{align}
\int  p(\boldsymbol\beta_k)\ln  \beta_k d{\boldsymbol\beta_k}\; \sim  \int  p(\boldsymbol\beta_{k-1}) \ln  \beta_{k-1} d\boldsymbol\beta_{k-1},
\end{align}
where we recall that the notation $\sim$ implies equality, except for an irrelevant additive constant. 
If we repeat this procedure  recursively  we get (\ref{eq:Sn1}).

Next consider  terms of the form 
\begin{align}\label{2entropypartial2}
\int p(\boldsymbol \beta_{k}) \ln  f(\beta_{k}|\beta_{k-1}) d\boldsymbol \beta_k.
\end{align}
Using (\ref{scale2}), we  have
\begin{align}\label{3entropypartial2}
  \int p(\boldsymbol\beta_k)\ln f(\beta_{k}|\beta_{k-1})d{\boldsymbol\beta_k}
 &=  \int p(\boldsymbol\beta_{k})\ln \left[\frac{1}{\beta_{k-1}}g_k\left(\frac{\beta_k}{\beta_{k-1}}\right)\right] d\boldsymbol\beta_k \cr
 &=  \int g(u)\ln g_k(u)d u  -\int p(\boldsymbol\beta_{k-1})\ln \beta_{k-1}d\boldsymbol\beta_{k-1}\cr
 &= B_k -\int p(\boldsymbol\beta_{k-1})\ln \beta_{k-1}d\boldsymbol\beta_{k-1},
\end{align}
where $B_k=\int_0^\infty   g_{k}(u) \ln g_k(u)  \, du$. Neglecting this additive constant we can  then write
\begin{align}\label{eq:Sn5}
&\int p(\boldsymbol\beta_k)\ln f(\beta_{k}|\beta_{k-1}) d{\boldsymbol\beta_k}\; \sim -  \int p(\boldsymbol\beta_{k-1}) \ln \beta_{k-1} d\boldsymbol\beta_{k-1}, 
\end{align}
which in view of (\ref{eq:Sn1}) yields    (\ref{eq:Sn3}), as desired.

\section{Derivation of  (\ref{H-gamma}) and  (\ref{H-inv-gamma})}

Here we calculate $p(\beta_N)$ explicitly in terms of Fox $H$-functions. We begin by introducing the variable
\be 
y=\frac{\beta _N}{\beta _0}=\prod_{j=1}^N\xi _j,
\ee
where $\xi _j=\beta _j/\beta _{j-1}$, so that $p(\beta_N)=g(y)/\beta_0$ and
\be 
g(y)=\int_0^\infty\cdots\int_0^\infty \prod_{j=1}^Ng_j(\xi _j)d\xi _j\delta (y-\xi _1\xi_2 \cdots \xi
_N).
\label{g}
\ee 
For $r>0$ we obtain  from (\ref{gengammaj}) that
\be 
g_j(\xi _j)=\frac{r\alpha _j^{\alpha _j}}{\Gamma (\alpha _j)}\xi _j^{r\alpha
_j-1}e^{-\alpha _j\xi _j^r},
\label{eq:gamma2}
\ee 
whilst for $r<0$ we find
\be 
g_j(\xi _j)=\frac{r'\alpha _j^{\alpha _j}}{\Gamma (\alpha _j)}\xi _j^{-r'\alpha
_j-1}e^{-\alpha _j\xi _j^{-r'}},
\label{eq:invgamma2}
\ee 
where we defined $r'=-r>0$.

Now applying the Mellin transform, defined as
\be 
{\cal M}[g;s] \equiv \int_0^\infty dy y^{s-1}g(y),
\ee 
to both sides of (\ref{g}), we find
\be 
{\cal M}[g;s] =\prod_{j=1}^N{\cal M}[g_j;s],
\label{Mellin_g}
\ee 
where 
\be 
{\cal M}[g_j;s] =\frac{\Gamma (\alpha _j+(s-1)/r)}{%
\alpha_j^{(s-1)/r}\Gamma (\alpha _j)}\label{Mellin-gamma}
\ee 
is the Mellin transform of (\ref{eq:gamma2}) and
\be 
{\cal M}[g_j;s] =\alpha _j^{(s-1)/r'}\frac{\Gamma
(\alpha _j+(1-s)/r')}{\Gamma (\alpha _j)}\label{Mellin-invgamma}
\ee 
is the Mellin transform of (\ref{eq:invgamma2}). Next, we use the following property of the Fox $H$-function \citep{Hfunction}. If the Mellin transform of $g(y)$ is
\[
{\cal M}[g;s]=\frac{\lambda ^{-s}\prod_{j=1}^m\Gamma
(b_j+B_j s)\prod_{j=1}^n\Gamma (1-a_j-A_j s)}{\prod_{j=m+1}^q\Gamma
(1-b_j-B_j s)\prod_{j=n+1}^p\Gamma (a_j+A_j s)}
\]
then
\be 
g(y)=H_{p,q}^{m,n}\left( 
\begin{array}{l}
(\boldsymbol a, \boldsymbol A) \\ 
(\boldsymbol b, \boldsymbol B)
\end{array}
\bigg |\lambda y\right) \label{Mellin-H},
\ee 
where we introduced the notation $(\boldsymbol x, \boldsymbol X)\equiv \{(x_1,X_1),\ldots ,(x_d,X_d)\}$, with $d\in \{p,q\}$.  Using (\ref{Mellin_g}),  (\ref{Mellin-gamma}) and (\ref{Mellin-H}) we obtain (\ref{H-gamma}), while using (\ref{Mellin_g}),  (\ref{Mellin-invgamma}) and (\ref{Mellin-H}) we get (\ref{H-inv-gamma}), as desired.

\section{Derivation of  (\ref{HgammaX}) and  (\ref{Hinv_gammaX})}

We start by considering the Laplace transform of the Fox $H$-function \citep{Hfunction}
\be 
\int_0^{\infty} dx\; x^{\gamma}e^{-sx}H_{p,q}^{m,n}\left( 
\begin{array}{l}
(\boldsymbol a, \boldsymbol A) \\ 
(\boldsymbol b, \boldsymbol B)
\end{array}
\bigg |\lambda x\right)=
s^{-(\gamma+1)}H_{p+1,q}^{m,n+1}\left( 
\begin{array}{l}
(\boldsymbol a, \boldsymbol A),(-\gamma,1) \\ 
(\boldsymbol b, \boldsymbol B)
\end{array}
\bigg |\lambda s^{-1}\right) 
\label{Laplace-H},
\ee 
where $(\boldsymbol x, \boldsymbol X)\equiv \{(x_1,X_1),\ldots ,(x_d,X_d)\}$, with $d\in \{p,q\}$. Using the identities

\be 
H_{p,q}^{m,n}\left( 
\begin{array}{l}
(\boldsymbol a, \boldsymbol A) \\ 
(\boldsymbol b, \boldsymbol B)
\end{array}
\bigg |z\right)=H_{q,p}^{n,m}\left( 
\begin{array}{l}
({\bf 1}-\boldsymbol b, \boldsymbol B)\\
({\bf 1}-\boldsymbol a, \boldsymbol A) 
\end{array}
\bigg |\frac{1}{z}\right)
\ee 
and
\be 
z^{\sigma}H_{p,q}^{m,n}\left( 
\begin{array}{l}
(\boldsymbol a, \boldsymbol A) \\ 
(\boldsymbol b, \boldsymbol B)
\end{array}
\bigg |z\right)=H_{p,q}^{m,n}\left( 
\begin{array}{l}
(\boldsymbol a+\sigma\boldsymbol A, \boldsymbol A)\\
(\boldsymbol b+\sigma\boldsymbol B, \boldsymbol B) 
\end{array}
\bigg |z\right)
\ee 
we may rewrite (\ref{Laplace-H}) as
\be 
\int_0^{\infty} dx x^{\gamma}e^{-sx}H_{p,q}^{m,n}\left( 
\begin{array}{l}
(\boldsymbol a, \boldsymbol A) \\ 
(\boldsymbol b, \boldsymbol B)
\end{array}
\bigg |\lambda x\right)=\frac{1}{\lambda^{\gamma +1}}
H_{q,p+1}^{n+1,m}\left( 
\begin{array}{l}
({\bf 1}-\boldsymbol b-(\gamma+1)\boldsymbol B, \boldsymbol B) \\ 
({\bf 1}-\boldsymbol a-(\gamma+1)\boldsymbol A, \boldsymbol A),(0,1)
\end{array}
\bigg |\frac{s}{\lambda}\right) 
\label{Laplace2-H}.
\ee

\noindent 
We are now in position to calculate the Laplace transform of $p(\beta_N )$. Using (\ref{H-gamma}) and  (\ref{Laplace2-H}), we get for the case $r>0$:
\begin{equation}
\label{Lap-H-gamma}
   \int_0^{\infty}d\beta_N\;\beta_N^{\gamma}e^{-\beta_N E} p(\beta_N )=
\frac{\beta_0^\gamma}{\omega_\rho^\gamma \Gamma(\boldsymbol\alpha)}    H_{ N,1 } ^{ 1,N }  \left( 
\begin{array}{c}
{((1-\gamma\rho){\bf 1}-\boldsymbol\alpha,\rho {\bf 1})} \\ 
{(0,1 )}
\end{array}
\bigg |\frac{\beta_0 E(\boldsymbol q)}{\omega_\rho }  \right).
\end{equation}
Similarly, in view of (\ref{H-inv-gamma}), the result for $r<0$ is
\begin{equation}
\label{Lap-H-inv-gamma}
    \int_0^{\infty}d\beta_N\;\beta_N^{\gamma}e^{-\beta_N E}p(\beta_N )=
\frac{(\beta_0\omega_\rho)^\gamma}{\Gamma(\boldsymbol\alpha)}    H_{ 0,N+1 } ^{ N+1,0 }  \left( 
\begin{array}{c}
{-}\\
{ (\boldsymbol\alpha-\gamma\rho {\bf 1},\rho{\bf 1}),(0,1)} 
\end{array}
\bigg |\omega_\rho\beta_0 E(\boldsymbol q)  \right).
\end{equation}
Using (\ref{Lap-H-gamma}) and (\ref{Lap-H-inv-gamma}) we obtain (\ref{HgammaX}) and  (\ref{Hinv_gammaX}) respectively.


\end{document}